\documentstyle[12pt,a41,epsfig]{article}

\begin{document}
\vskip 0.2cm
\hfill{YITP-SB-03-24}
\vskip 0.2cm
\hfill{INLO-PUB-08/03}\\[0.5cm]
\vskip 0.2cm
\centerline{\large\bf 
{ 
Differential Cross Sections for Higgs Production
} 
}
\vskip 0.4cm
\centerline {\sc V. Ravindran
}
\centerline{\it Harish-Chandra Research Institute,}
\centerline{\it Chhatnag Road, Jhunsi,}
\centerline{\it Allahabad, 211019, India.}
\vskip 0.2cm
\centerline {\sc J. Smith 
}
\centerline{\it C.N. Yang Institute for Theoretical Physics,}
\centerline{\it State University of New York at Stony Brook,
New York 11794-3840, USA.}
\vskip 0.2cm
\centerline {\sc W.L. van Neerven}
\centerline{\it Instituut-Lorentz}
\centerline{\it University of Leiden,}
\centerline{\it PO Box 9506, 2300 RA Leiden,}
\centerline{\it The Netherlands.}
\vskip 0.2cm
\vskip 0.2cm
\centerline{\bf Abstract}
\vskip 0.3cm
We review recent theoretical progress in evaluating higher
order QCD corrections to Higgs boson differential 
distributions at hadron-hadron colliders.

\section{Introduction}
The origin of the spontaneous symmetry breaking mechanism (ssbm) 
in particle physics is still unknown.  
In the Standard Model (SM) \cite{gsw} a vacuum expectation 
value (vev) $v=246$ GeV is given to components of a complex 
scalar doublet. Three of these fields generate mass terms for the 
$W^{\pm}$ and $Z$ bosons.  The remaining scalar is called the 
Higgs \cite{higgs} and has not been observed.  The present lower 
limit for $m_H$ from the LEP experiments \cite{lep} is 114 GeV/$c^2$.
In the SM the interaction vertices with scalar fields
are expressed in terms of $v$ and $m_H$. Extensions of the standard model to
incorporate supersymmetry contain several Higgs particles, both scalar and
pseudoscalar \cite{ghkd}.  In the minimal supersymmetric extension of the
Standard Model (MSSM), there are two Higgs doublets, so it is 
called the Two-Higgs-Doublet Model (2HDM). After implementing the
ssbm there are five Higgs bosons usually denoted by $h,H$, (CP even), 
$A$ (CP odd) and $H^{\pm}$.  At tree level their couplings and masses 
depend on two parameters, $m_A$ and the ratio of two vevs, 
parametrized by $\tan \beta = v_2/v_1$.  Since no Higgs particle 
has been found the region $m_A<$ 92 GeV/$c^2$ and $0.5 < \tan \beta < 2.4$ 
is experimentally excluded \cite{CERN}. For recent reviews on the 
theoretical status of total Higgs boson cross sections see \cite{review}.
Let us denote the neutral bosons h,H,A collectively by B. It is possible 
that a B will be detected at the Fermilab Tevatron $p -\bar p$ collider 
($\sqrt S =2$ TeV). If not it should hopefully be found at the Large Hadron 
Collider (LHC), a $p-p$ machine under construction at CERN 
($\sqrt S =14$ TeV).  If no B is found then the ssbm must be realized 
in a different way, possibly via dynamical interactions between the 
gauge bosons.  In parallel with the ongoing experimental effort 
higher order quantum chromodynamic (QCD) corrections to B production 
differential distributions are needed.  The leading order (LO) 
partonic reactions were calculated quite a long time ago and the 
next-to-leading order (NLO) corrections to these distributions
only recently. Note that the detection of the B via its
decay products depends on $m_B$.  The present mass limits allow the
decays $B\rightarrow \gamma\gamma$, $B\rightarrow b\bar b$ and
$B\rightarrow W W^*, Z Z^*$ where $W^*$ and $Z^*$ are virtual
vector bosons, which are detected as leptons and/or hadrons (jets).
As $m_B$ increases other decay channels 
(for example $B\rightarrow W^+W^-, ZZ$, $B\rightarrow t \bar t$),
open up requiring different experimental triggers. 
Higher order QCD corrections to these decay rates 
have also been calculated but will not be discussed here.  



\section{Higgs differential distributions}

For inclusive B-production one calculates the differential cross section in
the B transverse momentum ($p_T$) and its rapidity ($y$), 
which are functions of $m_B$, the partonic momentum fractions 
$x_1$, $x_2$ in the hadron beams,
and the partonic cm energy $\sqrt{s} =\sqrt{x_1 x_2 S}$. All other 
final state particles are integrated over. In contrast exclusive B-production
retains the information on these other particles.
In the SM the H couples to the gluons via quark loops with
the $\bar q q$H vertex proportional to $m_q$, so 
the t-quark loop is the most important.  
In the 2HDM the $gg$A amplitude with quark loops depends on both the
masses of the quarks and $\beta$.  In LO the $g + g \rightarrow H$
cross section (order $\alpha_s^2$) containing the top-quark triangle graph, 
was computed in \cite{wil}. However here $p_T=0$, so we need   
a two-to-two body partonic process to produce a Higgs with a finite $p_T$.
Note that these are NLO processes with respect to the 
total B production cross section and of order $\alpha_s^3$.  
At small $x$ the gluon density $g(x,Q^2) > q_i(x,Q^2) > \bar q_i(x,Q^2)$, 
($q_i$ stand for $u,d,s,c,b,t$ quarks) so we expect that, in order
of importance, the dominant production channels are
\begin{eqnarray}
\label{eqn1}
\qquad g + g && \rightarrow g + B ,\nonumber\\[2ex]  
\qquad g + q_i(\bar q_i) && \rightarrow q_i(\bar q_i) + B , \nonumber\\[2ex] 
\qquad q_i + \bar q_i && \rightarrow g + B  \,. 
\end{eqnarray}
These are the LO order Born reactions for Higgs $p_T$ and $y$ distributions.
The corresponding Feynman diagrams contain heavy quark box graphs.
The B differential distributions for the reactions in Eq. (\ref{eqn1})
were computed for B=H in \cite{ehsb} and for B=A in \cite{kao}.
The total cross section, which also contains the virtual QCD corrections
to the $gg$B top-quark triangle, was calculated in \cite{dawson}, \cite{sdgz1}
and \cite{sdgz2}.
The expressions in \cite{sdgz2} for the two-loop 
graphs with finite $m_t$ and $m_B$ are very complicated. 
Furthermore also the
two-to-three body reactions (e.g. $g+g\rightarrow g+g+B$ involving
pentagon loops) have been computed in \cite{dkosz} using helicity methods. 
From these results it is clear that it will be very 
difficult to obtain the NLO (order $\alpha_s^4$) corrections
to the B differential distributions as functions of both $m_t$ and $m_B$.

Fortunately one can simplify the calculations if one takes the
limit $m_t \rightarrow \infty$. In this case the Feynman graphs are obtained 
from an effective Lagrangian describing the direct $Bgg$ coupling. 
An analysis
in \cite{kls} in NLO reveals that the error introduced by taking the
$m_t\rightarrow \infty$ limit is less than about $5\%$ provided $m_B\le 2~m_t$.
The two-to-three body processes were computed with the effective
Lagrangian approach for the B in \cite{kdr} and \cite{kade} respectively 
using helicity methods.
The one-loop corrections to the two-to-two body reactions above were 
computed for the H in \cite{schmidt} and the A in \cite {kao}.
These NLO matrix elements (order $\alpha_s^4$)
were used to compute the $p_T$ and $y$
distributions of the H in \cite{fgk}, \cite{rasm1}, \cite{glosser}, 
\cite{glsc} and the A in \cite{fism}. For ways to differentiate
between production of H and A see \cite{bjf} and references therein.

In the large $m_t$ limit the Feynman rules (see e.g. \cite{kdr})
for scalar H production can be derived from the following
effective Lagrangian density
\begin{eqnarray}
\label{eqn2}
{\cal L}^{\rm H}_{eff}=G_{\rm H}\,\Phi^{\rm H}(x)\,O(x) \quad
\mbox{with} \quad O(x)=-\frac{1}{4}\,G_{\mu\nu}^a(x)\,G^{a,\mu\nu}(x)\,,
\end{eqnarray}
whereas pseudoscalar A production is obtained from
\begin{eqnarray}
\label{eqn3}
&&{\cal L}_{eff}^{\rm A}=\Phi^{\rm A}(x)\Big (G_{\rm A}\,O_1(x)+
\tilde G_{\rm A}\,O_2(x)\Big )\,, \quad \quad \mbox{with} 
\nonumber\\[2ex]
&&O_1(x)=-\frac{1}{8}\,\epsilon_{\mu\nu\lambda\sigma}\,G_a^{\mu\nu}(x)\,
G_a^{\lambda\sigma}(x) \quad ,\quad
O_2(x) =-\frac{1}{2}\,\partial^{\mu}\,\sum_{i=1}^{n_f}
\bar q_i(x)\,\gamma_{\mu}\,\gamma_5\,q_i(x)\,,
\end{eqnarray}
where $\Phi^{\rm H}(x)$ and  $\Phi^{\rm A}(x)$ represent the scalar and
pseudoscalar fields respectively and $n_f$ denotes the number of light
flavours.
Furthermore the gluon field strength tensor is given by $G_a^{\mu\nu}(x)$ 
($a$ is the colour index) and the quark field is denoted by $q_i(x)$.
The factors in the definitions of $O$,$O_1$ and $O_2$ are chosen in such 
a way that the vertices are normalised to the effective coupling 
constants $G_{\rm H}$,
$G_{\rm A}$ and $\tilde G_{\rm A}$. 
The latter are determined by the
triangular loop graphs, which describe the decay processes
${\rm B} \rightarrow g + g$ with ${\rm B}={\rm H},{\rm A}$, including all 
QCD corrections and taken in the limit $m_t\rightarrow \infty$, namely
\begin{eqnarray}
\label{eqn4}
G_{\rm B}&=&-2^{5/4}\,a_s(\mu_r^2)\,G_F^{1/2}\,
\tau_{\rm B}\,F_{\rm B}(\tau_{\rm B})\,{\cal C}_{\rm B}
\left (a_s(\mu_r^2),\frac{\mu_r^2}{m_t^2}\right )\,,
\nonumber\\[2ex]
\tilde G_{\rm A}&=&-\Bigg [a_s(\mu_r^2)\,C_F\,\left (\frac{3}{2}-3\,
\ln \frac{\mu_r^2}{m_t^2}\right )+\cdots \Bigg ]\,G_{\rm A}\,,
\end{eqnarray}
where $a_s(\mu_r^2)$ is defined by
\begin{eqnarray}
\label{eqn5}
a_s(\mu_r^2)=\frac{\alpha_s(\mu_r^2)}{4\pi}\,,
\end{eqnarray}
with $\alpha_s(\mu_r^2)$ the running coupling constant and $\mu_r$ the 
renormalization scale. 
Further $G_F$ represents the Fermi constant and the
functions $F_{\rm B}$ are given by
\begin{eqnarray}
\label{eqn6}
&& F_{\rm H}(\tau)=1+(1-\tau)\,f(\tau)\,, \qquad  F_{\rm A}(\tau)
=f(\tau)\,\cot \beta\,,\qquad
\tau=\frac{4\,m_t^2}{m_B^2} \,,
\nonumber\\[2ex]
&&f(\tau)=\arcsin^2 \frac{1}{\sqrt\tau}\,, \quad \mbox{for} \quad \tau \ge 1\,,
\nonumber\\[2ex]
&& f(\tau)=-\frac{1}{4}\left ( \ln \frac{1-\sqrt{1-\tau}}{1+\sqrt{1-\tau}}
+\pi\,i\right )^2 \quad \mbox{for} \quad \tau < 1\,,
\end{eqnarray}
where $\beta$ denotes the mixing angle in the 2HDM.
In the large $m_t$-limit
we have
\begin{eqnarray}
\label{eqn7}
 \mathop{\mbox{lim}}\limits_{\vphantom{\frac{A}{A}} \tau \rightarrow \infty}
F_{\rm H}(\tau)=\frac{2}{3\,\tau}\,, \qquad
 \mathop{\mbox{lim}}\limits_{\vphantom{\frac{A}{A}} \tau \rightarrow \infty}
F_{\rm A}(\tau)=\frac{1}{\tau}\,\cot \beta\,.
\end{eqnarray}
The coefficient functions ${\cal C}_{\rm B}$ originate from the corrections
to the top-quark triangle graph provided one takes the
limit $m_t\rightarrow \infty$.  The coefficient functions were computed
up to order $\alpha_s^2$ in \cite{kls}, \cite{cks} for the H
and in \cite{cksb} for the A. The answer depends upon
$a_s^{(5)}$, the running coupling in the five-flavour number scheme,
and we only need the first order terms 
\begin{eqnarray}
\label{eqn8}
&&{\cal C}_H\Big(a_s(\mu_r^2),\frac{\mu_r^2}{m_t^2}\Big)=
1 +  11 a_s^{(5)} + ..., \nonumber \\ 
&&{\cal C}_A\Big(a_s(\mu_r^2),\frac{\mu_r^2}{m_t^2}\Big)=1.
\end{eqnarray}
The last result holds to all orders because of the
Adler-Bardeen theorem \cite{adba}. 
%

\section{Numerical results at moderate $p_T$}

The hadronic cross section $d\sigma$ for
$H_1(P_1)+H_2(P_2) \rightarrow B(-P_5) + X$ is obtained from the partonic 
cross sections $d\sigma_{ab}$ for the reactions in Eq. (\ref{eqn1})
and their NLO corrections
[for example $g(p_1)+g(p_2) \rightarrow g(-p_3) + g(-p_4) + B(-p_5)$]
using 
\begin{eqnarray}
\label{eqn9}
S^2 \frac{d^2~\sigma^{{\rm H_1H_2}}}{d~T~d~U}(S,T,U,m_B^2)&=& \sum_{a,b=q,g}
\int_{x_{1,{\rm min}}}^1 \frac{dx_1}{x_1} \int_{x_{2,{\rm min}}}^1
\frac{dx_2}{x_2}\,
f_a^{\rm H_1}(x_1,\mu^2)
\nonumber\\[2ex]
&&\times f_b^{\rm H_2}(x_2,\mu^2)\,s^2
\frac{d^2~\sigma_{ab}}{d~t~d~u} (s,t,u,m_B^2,\mu^2)\,.
\end{eqnarray}

Here $f_a^H(x,\mu^2)$ is the parton density for parton $a$ in hadron
$H$ at factorization/renormalization scale $\mu$ and the hadronic 
kinematical variables are defined by
\begin{eqnarray}
\label{eqn10}
S=(P_1+P_2)^2 \,, \qquad T=(P_1+P_5)^2\,, \qquad U=(P_2+P_5)^2 \,.
\end{eqnarray}
\begin{figure}[th]
\centerline{\psfig{file=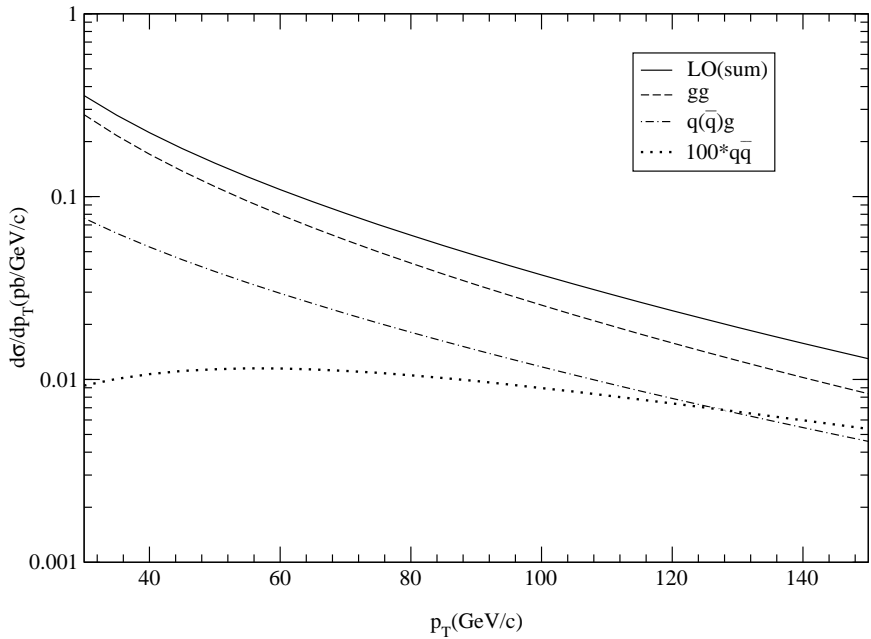,angle=270, width=4.5in}}
\vspace*{10pt}
\caption{
The differential cross section $d~\sigma/dp_T$
integrated over the whole rapidity range (see Eq. (\ref{eqn13}))
with $m_H=120~{\rm GeV/c^2}$ and $\mu^2=m_H^2+p_T^2$.
The LO plots are presented for the subprocesses $gg$ (long-dashed line),
$q(\bar q)g$ (dot-dashed line)
and  $100 \times (q\bar q)$ (dotted line)
using the parton density set MRST98(lo05a.dat).
}
\label{fig:fig1}
\end{figure}

The latter two invariants can be expressed in terms of the
$p_T$ and $y$ variables by 
\begin{eqnarray}
\label{eqn11}
&&T=m_B^2-\sqrt{S}\sqrt{p_T^2+m_B^2}\cosh y 
        + \sqrt{S}\sqrt{p_T^2+m_B^2}\sinh y\,,
\nonumber \\ 
&&U=m_B^2-\sqrt{S}\sqrt{p_T^2+m_B^2}\cosh y 
        - \sqrt{S}\sqrt{p_T^2+m_B^2}\sinh y\,.
\end{eqnarray}
In the case parton $p_1$ emerges from hadron $H_1(P_1)$ and parton
$p_2$ emerges from hadron $H_2(P_2)$ we can establish the following relations
\begin{eqnarray}
\label{eqn12}
&& p_1=x_1\,P_1\,, \qquad p_2=x_2\,P_2 \,,
\nonumber\\[2ex]
&& s=x_1\,x_2\,S \,,\quad t=x_1(T-m_B^2)+m_B^2 \,,\quad u=x_2(U-m_B^2)+m_B^2\,,
\nonumber\\[2ex]
&& x_{1,{\rm min}}=\frac{-U}{S+T-m_B^2}\,, \qquad
x_{2,{\rm min}}=\frac{-x_1(T-m_B^2)-m_B^2}{x_1S+U-m_B^2}\,.
\end{eqnarray}

\begin{figure}[th]
\centerline{\psfig{file=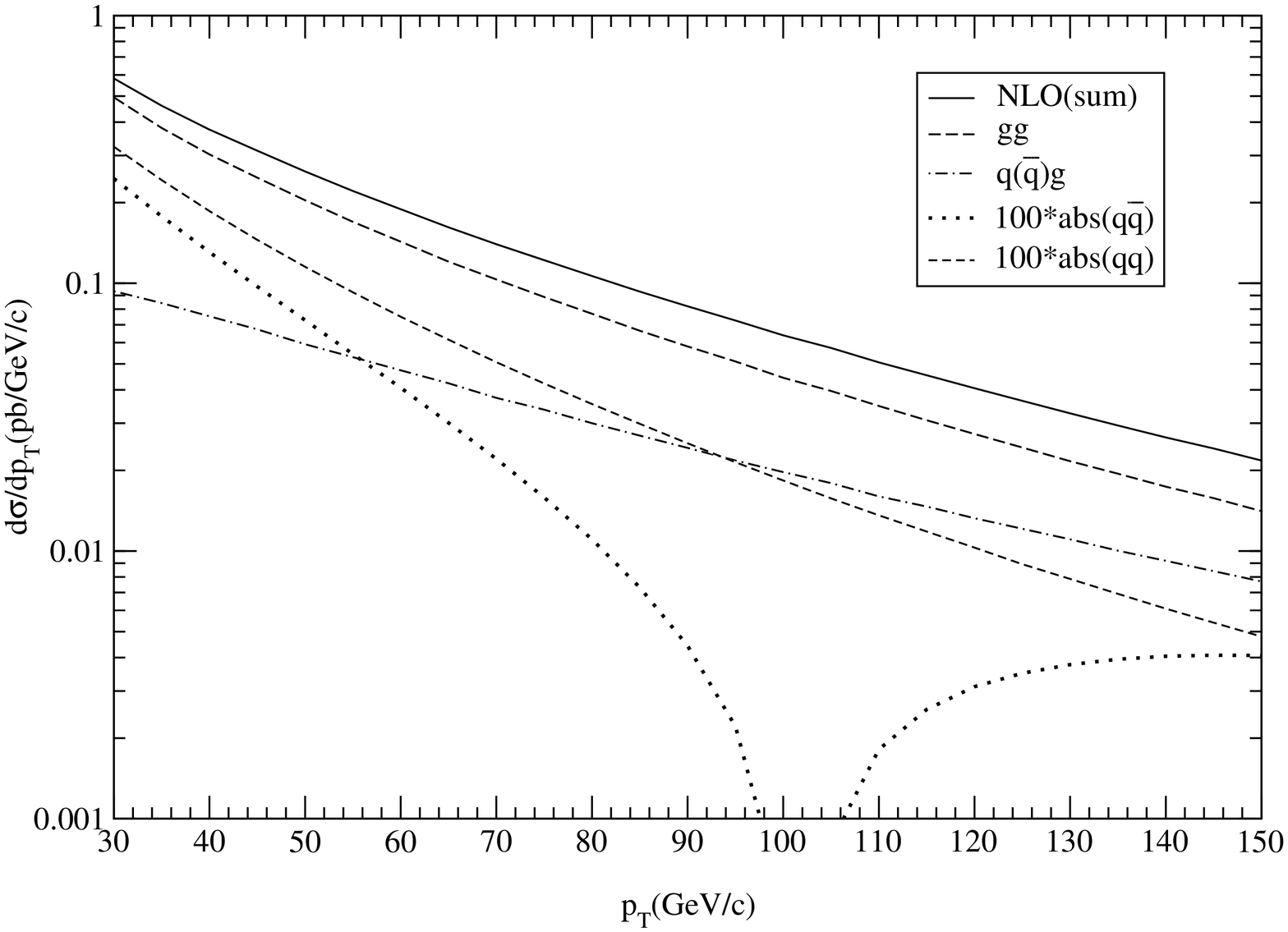,angle=270, width=4.5in}}
\vspace*{10pt}
\caption{
Same as Fig. 1 in NLO except for $100*{\rm abs}(q\bar q)$
(dotted line) and
the additional subprocess $100 \times {\rm abs}(qq)$ (short-dashed line)
using the parton density set MRST98(ft08a.dat).
}
\label{fig:fig2}
\end{figure}
Since the differential cross section contains terms in 
$m_B/p_T \ln^i(m_B/p_T)$, $(i=1,2,3)$ which are not integrable at $p_T=0$,
we cannot integrate over $p_T$ down to $p_T=0$ to find the $y$-distribution.
However we can integrate over $y$ to find the $p_T$-distribution, 
valid over a range in $p_T$ where there are no large logarithms,
(say $p_T > 30$ ${\rm GeV/c}$),
\begin{eqnarray}
\label{eqn13}
\frac{d~\sigma^{{\rm H_1H_2}}}{d~p_T}(S,p_T^2,m_B^2)=
\int_{-y_{{\rm max}}}^{y_{{\rm max}}} dy
\,\frac{d^2~\sigma^{{\rm H_1H_2}}}{d~p_T~d~y}
(S,p_T^2,y,m_B^2)\,,
\end{eqnarray}
with  a fixed $y_{{\rm max}}$. 
The calculation of the NLO B differential distributions
requires the virtual corrections to the reactions in Eq. (\ref{eqn1}) 
and the NLO two-to-three body reactions (order $\alpha_s^4$). 
Hence one needs a regularization scheme, and renormalization 
and mass factorization (which introduces the scales $\mu_r$
and $\mu_f$ respectively). The presence of the $\gamma_5$ matrix
in the pseudoscalar contributions makes everything even more complicated
and we refer to \cite{fism} for details.
In \cite{fgk} helicity amplitudes were used and the fully exclusive 
two-to-three body reactions were calculated numerically.  Very few formulae 
were presented.  In \cite{rasm1} the cancellations of the UV and IR 
singularities were done algebraically leading to 
the H inclusive distributions. 
Many analytical results were given but the
two-to-three body matrix elements were too long to publish.
In \cite{glosser}, \cite{glsc} helicity
amplitudes were used for the H inclusive calculation and complete analytical
results were provided. The numerical results from these three
papers have been compared against each other and they agree.

\begin{figure}[th]
\centerline{\psfig{file=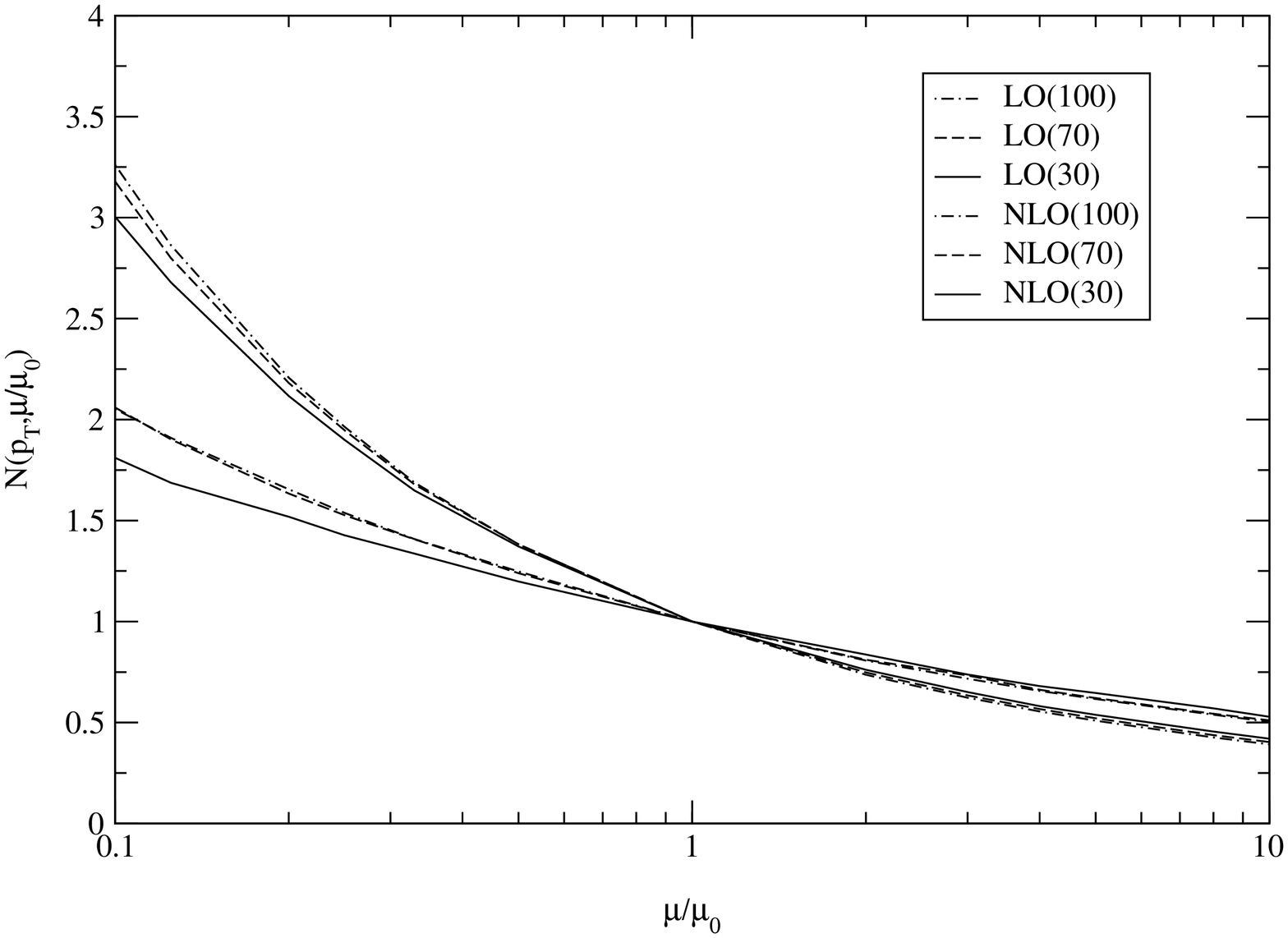,angle=270, width=4.5in}}
\vspace*{10pt}
\caption{
The quantity $N(p_T,\mu/\mu_0)$ (see Eq. (\ref{eqn14})), plotted in the range
$0.1<\mu/\mu_0<10$ at fixed values of $p_T$
with $m_H=120~{\rm GeV/c^2}$ and $\mu_0^2=m_H^2+p_T^2$
using the MRST98 parton density sets.
The results are shown for
$p_T=30~{\rm GeV/c}$ (solid line), $p_T=70~{\rm GeV/c}$ (dashed line),
$p_T=100~{\rm GeV/c}$ (dot-dashed line).
The upper three curves on the left hand side are the LO results whereas
the lower three curves refer to NLO.
}
\label{fig:fig3}
\end{figure}
We put $n_f=5$ in $a_s(\mu_r^2)$, $\sigma_{ab}$ and  
$f_a^H(x,\mu^2)$ in Eq.(\ref{eqn9}).  For simplicity $\mu_r=\mu_f=\mu$  
and we take $\mu^2=m_H^2+p_T^2$ for our plots.
Further we have used the parton density 
sets MRST98 \cite{mrst98}, MRST99 \cite{mrst99}, GRV98 \cite{grv}, 
CTEQ4 \cite{cteq4} and CTEQ5 \cite{cteq5}. 

We want to emphasize that the magnitudes of the 
cross sections are extremely sensitive to the choice of the 
renormalization scale because the effective coupling constants 
in Eq.(\ref{eqn4}) are proportional to $\alpha_s(\mu_r)$,
which implies that $d\sigma^{\rm LO}\sim \alpha_s^3$ and
$d\sigma^{\rm NLO}\sim \alpha_s^4$. However the slopes 
of the differential distributions are less sensitive to the 
scale choice if they are only plotted over a limited range.
For the computation of the $gg$B effective coupling constants
in Eq. (\ref{eqn4}) 
we take $m_t=173.4~{\rm GeV/c^2}$ and 
$G_F=1.16639~{\rm GeV}^{-2}=4541.68~{\rm pb}$.
Here we will only give results for H production at the LHC.  
Lack of space limits us to showing only $p_T$-distributions.
The LO and NLO $p_T$ differential cross sections in Eq. (\ref{eqn13}) with 
$m_H=120$ GeV$/c^2$ are shown in Figs. 1 and 2 respectively. 
The MRST98 parton densities \cite{mrst98} were used for these plots. 
We note that the NLO results from the $q(\bar q) g$ and $qq$ channels 
are negative at small $p_T$ so we have plotted their absolute values 
multiplied by 100. Clearly the $gg$ reaction dominates. The $q(\bar q)g$
reaction is lower by a factor of about five.  

Regarding the corresponding distributions for $A$-production they have the
same shape dependence in LO and differ slightly in NLO. Therefore the
only difference is in the overall couplings in Eq. (\ref{eqn4}).
In particular if $\tan\beta=1$ then $d\sigma_A/d\sigma_H=9/4$ for all values
of $m_H$ and $m_A$ in LO. There are small differences from 9/4 in NLO. 
Plots are presented in \cite{fism}.  

\begin{figure}[th]
\centerline{\psfig{file=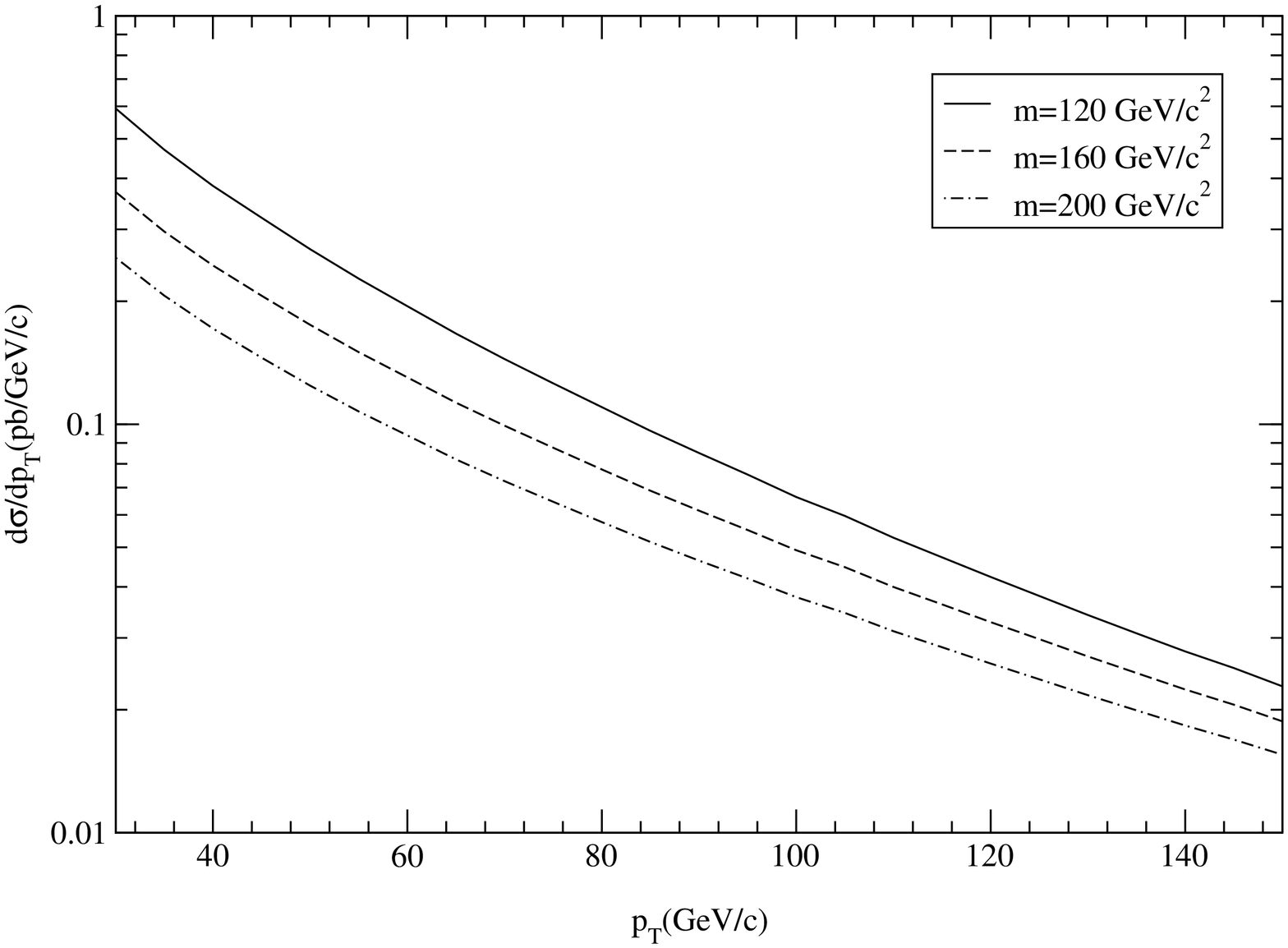,angle=270, width=4.5in}}
\vspace*{10pt}
\caption{
The mass dependence of $d~\sigma^{\rm NLO}/dp_T$
(see Eq. (\ref{eqn13})) using the set MRST99(cor01.dat)
 with $\mu^2=m_H^2+p_T^2$ for Higgs masses
$m_H=120~{\rm GeV/c^2}$ (solid line),
$m_H=160~{\rm GeV/c^2}$ (dashed line) and
$m_H=200~{\rm GeV/c^2}$ (dot-dashed line).
}
\label{fig:fig4}
\end{figure}

We now show the scale dependence of the distributions. 
We have chosen the scale factors 
$\mu=2\mu_0$, $\mu=\mu_0$ and $\mu=\mu_0/2$ with $\mu_0^2= m_H^2 + p_T^2$
and plot in Fig. 3 the quantity
\begin{eqnarray}
\label{eqn14}
N\left (p_T,\frac{\mu}{\mu_0}\right )
=\frac{d\sigma(p_T,\mu)/dp_T}{d\sigma(p_T,\mu_0)/dp_T} 
\end{eqnarray}
in the range $0.1 < \mu/\mu_0 < 10$ at fixed values of $p_T=$ 30, 70 and 
~{\rm 100 GeV/c}.
The upper set of curves at small $\mu/\mu_0$ are for LO and the
lower set are for NLO. 
Notice that the NLO plots at 70 and 100 are 
extremely close to each other and it is hard to distinguish between them.
One sees that the slopes of the LO curves are
larger that the slopes of the NLO curves. This is an indication that
there is a small improvement in stability in NLO, which was expected. 
However there is no sign of a flattening or an optimum in either of these
curves which implies that one will have to calculate the
differential cross sections in NNLO to find a better stability under
scale variations.

Next we show the mass dependence of the NLO result 
in Fig. 4 using the MRST99 parton densities. The differential distribution
drops by a factor of two as $m_H$ increases from 120 to 180 GeV/$c^2$.
    
There are two other uncertainties which affect the predictive power of the
theoretical cross sections. The first one concerns the rate of convergence
of the perturbation series which is indicated by the $K$-factor defined by
\begin{eqnarray}
\label{eqn15}
K=\frac{d~\sigma^{\rm NLO}}{d~\sigma^{\rm LO}} \,. 
\end{eqnarray}
Depending on the parton density set the $K$-factors
are pretty large and vary from 1.4 at $p_T=30~{\rm GeV/c}$ 
to 1.7 at $p_T=150~{\rm GeV/c}$ for both $H$ and $A$ production. 
Another uncertainty is the dependence of the $p_T$ distribution on the specific
choice of parton densities, which can be expressed by the factors like
\begin{eqnarray}
\label{eqn16}
R^{\rm CTEQ}=\frac{d~\sigma^{\rm CTEQ}}{d~\sigma^{\rm MRST}}\quad \,,\quad
R^{\rm GRV}=\frac{d~\sigma^{\rm GRV}}{d~\sigma^{\rm MRST}}\,,
\end{eqnarray}
and are generally above unity. Again these factors are essentially identical
for H and A production. For specific values consult \cite{fgk}, \cite{rasm1},
\cite{glsc} and \cite{fism}.

\begin{figure}[th]
\centerline{\psfig{file=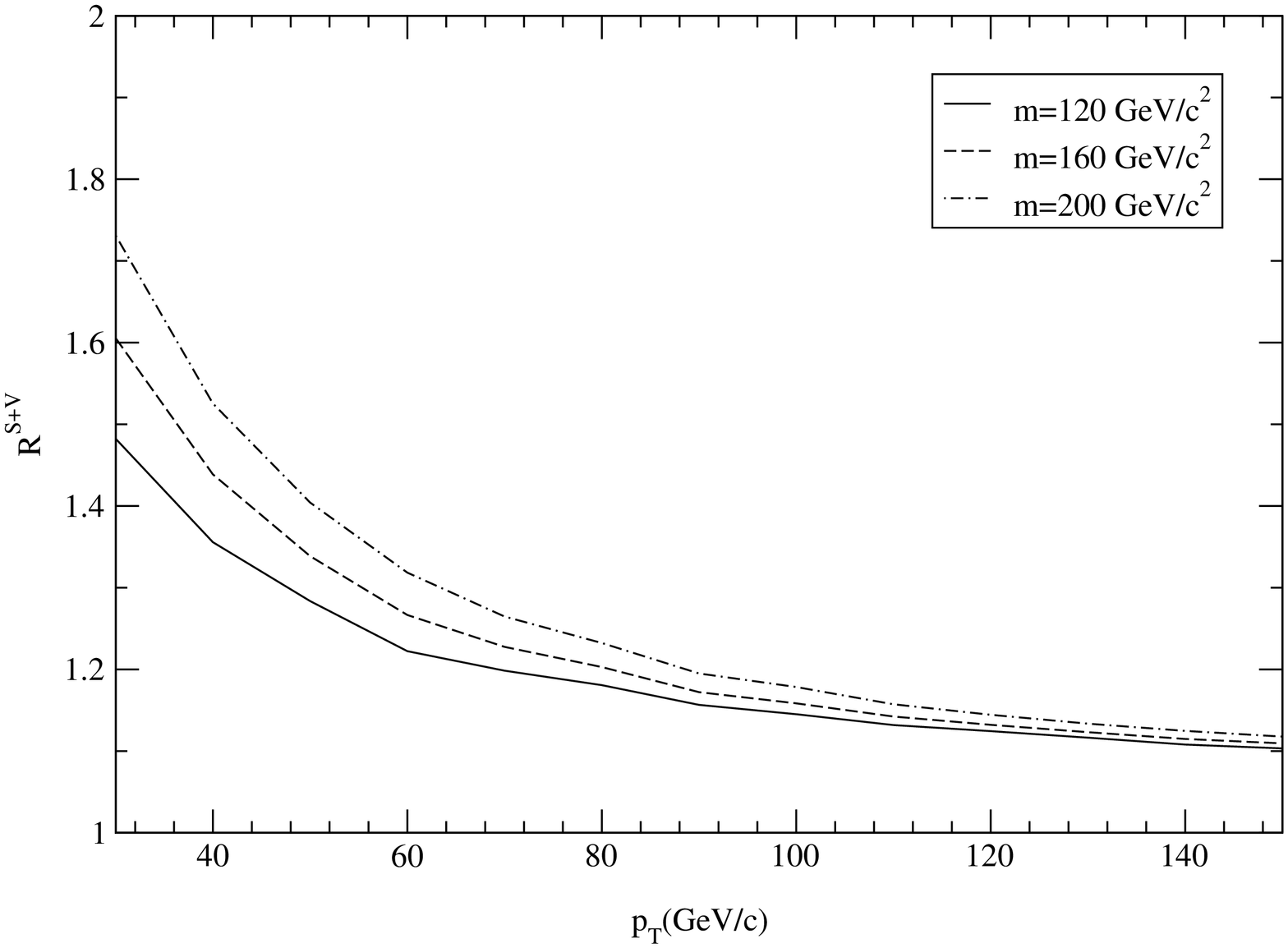,angle=270, width=4.5in}}
\vspace*{10pt}
\caption{
The ratio $R^{\rm S+V}$ in Eq. (\ref{eqn17}) for the $p_T$ distributions 
using the set MRST99(cor01.dat) with $\mu^2=m_H^2+p_{T,{\rm min}}^2$
and various Higgs masses given by
$m_H=120~{\rm GeV/c^2}$ (solid line),
$m_H=160~{\rm GeV/c^2}$ (dashed line) and
$m_H=200~{\rm GeV/c^2}$ (dot-dashed line).
}
\label{fig:fig5}
\end{figure}
The reason why the parton density sets yield different results for 
the $p_T$-distributions can be mainly attributed to the small 
$x$-behaviour of the gluon density because gluon-gluon fusion is the dominant
production mechanism. Future HERA data will have to provide us 
with unique gluon densities before we can make more accurate predictions 
for the Higgs differential distributions. 

\section{Numerical results at large $x$.}

Near threshold the longitudinal momentum fractions $x_i$ approach unity.
In this region the soft-plus-virtual (S+V) gluons and 
collinear $q-\bar q$ pairs dominate the NLO corrections
to the partonic cross sections.  The S+V gluon parts
of these cross sections are obtained by omitting the 
hard contributions which are regular at $s_4=(p_3+p_4)^2=0$
and adding the pieces from the virtual contributions.
These two contributions constitute the S+V gluon approximation. 
To study its validity we show in Fig. 5 the ratio
\begin{eqnarray}
\label{eqn17}
R^{\rm S+V}=\frac{d~\sigma^{\rm S+V}}{d~\sigma^{\rm EXACT}}\,,
\end{eqnarray}
for the NLO contributions to the $p_T$ distribution 
(here $p_{\rm T,min} = 30$ GeV/c).
One expects that the approximation becomes better 
at larger transverse momenta where $p_T$ approaches 
the boundary of phase space at $x=1$.
However in Fig. 5 the highest value of
$p_T$, given by $p_T=150~{\rm GeV/c}$, is still very small with respect to 
$p_{T,{\rm max}}\sim \sqrt{S}/2=7\times 10^{3}~{\rm GeV/c}$. 
Therefore it is rather fortuitous that the approximation works so well 
for $p_T>100~{\rm GeV/c}$ where one obtains $R^{\rm S+V}<1.2$.

\begin{figure}[th]
\centerline{\psfig{file=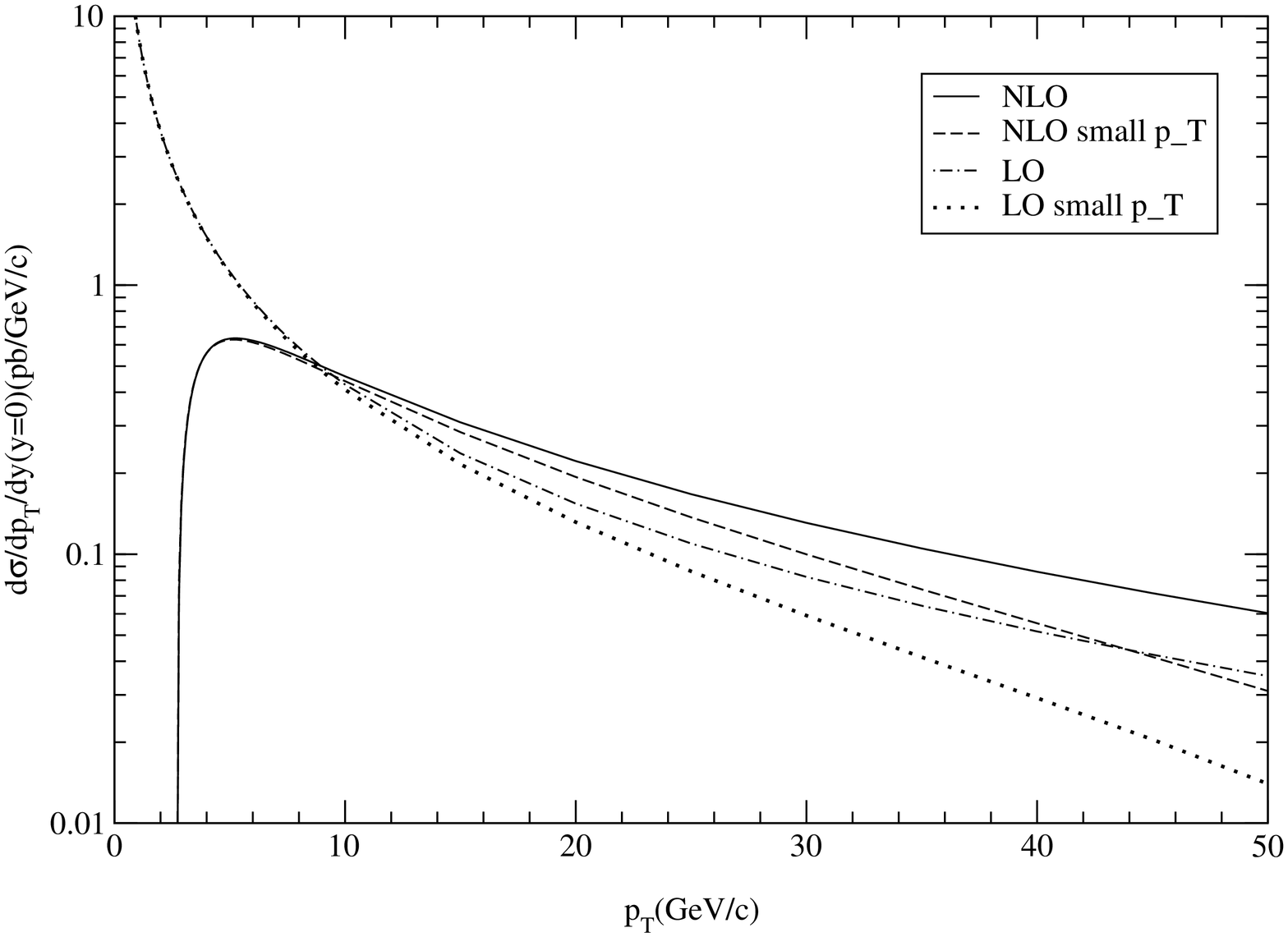,angle=270, width=4.5in}}
\vspace*{10pt}
\caption{
The Higgs $p_T$ spectrum compared to the small-$p_T$ limit formula
(see Eq. (\ref{eqn18})) at both LO and NLO with CTEQ5L and CTEQ5M1
parton densities respectively. All curves are calculated for
$y=0$ in the $m_t=\infty$ effective theory.
}
\label{fig:fig6}
\end{figure}
The S+V gluon approximation overestimates
the exact NLO result but the difference decreases when 
the $p_T$ increases. In particular for $p_T>200~{\rm GeV/c}$
the S+V approximation is good enough so that resummation 
techniques could be used to give a better estimate of the Higgs boson 
$p_T$ distribution corrected up to all orders in perturbation theory.
Note that the boundary of phase space is also approached
when $m_B$ increases at fixed $p_T$, but, according to the results in Fig. 5,  
the S+V approximation does not improve. 
In \cite{glsc} it was noted that increasing $m_B$ makes the
terms in $\ln(m_B/p_T)$ larger at fixed $p_T$. So the range in
$p_T$ where the small $p_T$-logarithms dominate increases with increasing
$m_B$. This leads us to the next topic.

\section{Numerical results at small $p_T$}

We noted previously that there are terms in $m_B/p_T\ln^i(m_B/p_T)$
which are dominant in the small $p_T$ region. In this region 
the double differential cross section can be expanded
as follows
\begin{eqnarray}
\label{eqn18}
\frac{d\sigma}{dp_T^2 dy}=
\frac{\sigma_0}{s}
\frac{m_B^2}{p_T^2}
\Big[ \sum_{m=1}^{2} \sum_{n=0}^{2m-1} \Big(\frac{\alpha_s}{2\pi}\Big)^m
C_{mn}\Big(\ln\frac{m_B^2}{p_T^2}\Big)^n + ... \Big]\,,
\end{eqnarray}
where the next order terms start with $m=3$ (order $\alpha_s^3$),
and $\sigma_0$ denotes one of the partonic cross sections for the 
LO processes in Eq.(\ref{eqn1}), which are order $\alpha_s^2$.
In \cite{glsc} the $C_{mn}$ were determined from their explicit NLO
results in terms of certain factors $A^{(1)}$, $A^{(2)}$,
$B^{(1)}$, $B^{(2)}$, $C^{(1)}_{gg}$, $C^{(1)}_{gq}$  which multiply
convolutions of splitting functions with parton densities. These factors
had been previously determined in \cite{arka}, \cite{cat1} from other
reactions, so it was gratifying to see the consistency between the 
different results.
In Fig. 6 we show a comparison of the inclusive result in Eq. (\ref{eqn9})
at both LO and NLO versus the corresponding result from the 
small $p_T$-limit formula in Eq. ({\ref{eqn18}). 
The latter works very well for $p_T$ smaller than 
10 GeV/c and agrees with the exact calculation for $p_T$ smaller than
5 GeV/c. The authors then propose another large $x$ approximation involving
both the S+V and small $p_T$ terms. We refer to their paper for details. 

Note that the resummation of the logarithms in Eq. (\ref{eqn18}) 
can be carried out following the procedure in \cite{css}. This leads to
a change in the shape of the $p_T$-spectrum at small $p_T$. Results
can be found in \cite{cpy}, \cite{fg}, \cite{fin}.
 
In conclusion we note that the NLO corrections to the B differential
distributions are now completely known and resummation methods have been
used to study the region near $p_T=0$.

Acknowledgement: We thank C. Glosser and C. Schmidt for giving us the 
inputs for Fig.6. We also thank B. Field, M. Tejeda-Yeomans, S. Dawson
R. Kauffman, D. de Florian, M. Grazzini and Z. Kunszt for 
discussions and comments.

\end{document}